\documentstyle[11pt,Cargesepasp,twoside,epsf,psfig]{article}
\def\edcomment#1{\iffalse\marginpar{\raggedright\sl#1\/}\else\relax\fi}
\reversemarginpar

\begin{document}
\title{The Structure of Prestellar Cores as derived from ISO Observations}
 \author{Aurore Bacmann}
\affil{Astrophysical Institute Jena, Schillerg\"a\ss chen 2--3, D-07745 Jena, Germany}
\author{Philippe Andr\'e}
\affil{Service d'Astrophysique, CEA/DSM/DAPNIA, CE-Saclay, 91191 Gif-Sur-Yvette Cedex, France}
\author{Derek Ward-Thompson}
\affil{Department of Physics \& Astronomy, Cardiff University, P.O. Box 913, 5 The Parade, Cardiff, UK}

\begin{abstract}

Observations of dark cloud cores have been carried out in the mid-infrared using ISOCAM and in the far-infrared using ISOPHOT, both aboard the Infrared Space Observatory. The cores are in most cases detected in emission at 200 and 170\,$\mu$m, remain undetected at 90\,$\mu$m and are seen in absorption against the diffuse mid-infrared background at 7\,$\mu$m. The observations are consistent with the cores being pre-stellar and not having a central heating source, and yield core temperatures of $\sim 11-13$\,K. We were able to determine the density structure of the cores up to radii that extend beyond the sensitivity limit of previous submillimetre studies. The column density profiles of the cores studied here flatten out in the centre, as shown by previous submillimetre continuum results, and interestingly, in a few cases, the profiles steepen with radius beyond $\sim 5000-10000$\,AU and present sharp edges. 
This could indicate that these cores are decoupled from their parent 
molecular cloud and represent finite reservoirs of mass 
for subsequent accretion. 
\end{abstract}

Keywords: pre-stellar cores; ISO; density profile; cloud structure; L1544

\section{Introduction}

Although our understanding of the star formation process has made progress in the past three decades, many aspects remain obscure. The study of the stages immediately preceding and following the onset of gravitational collapse are of particular interest since in some respect, they must govern the origins of stellar masses. According to present ideas on isolated star formation (Shu, Adams, \& Lizano 1987), the first evolutionary phase in the path from molecular cloud to main sequence star involves the formation
of gravitationally-bound, starless dense cores (e.g. Myers 1999). These pre-stellar cores (Ward-Thompson et al. 1994)
are thought to be supported against their own gravitation by thermal, and possibly magnetic/turbulent pressure. They evolve towards higher degrees of concentration through ambipolar diffusion (e.g. Mouschovias 1991) and/or the dissipation of turbulence (e.g. Nakano 1998) until some time as yet not well understood, when they become unstable and collapse dynamically to form a 
Class 0 protostar (Andr\'e, Ward-Thompson, \& Barsony 1993), which accretes 
mass from its surrounding envelope.
Recent studies have shown that the mass-accretion rate in the protostellar phase strongly depends on the density structure of the core in the pre-stellar phase (Foster \& Chevalier 1993, Henriksen, Andr\'e, \& Bontemps 1997). The outer core density profile plays a fundamental role by determining whether the mass
reservoir ultimately available for star formation is effectively
finite or infinite. 
Past (sub)millimetre continuum observations have shown that unlike protostellar envelopes which have a highly centrally peaked density profile 
($\rho\propto r^{-1.5}$ to $\rho\propto r^{-2}$ -- e.g. Ladd et al. 1991), 
pre-stellar cores have density profiles that flatten out near their centres 
(see Andr\'e, Ward-Thompson, Barsony 2000 for a review) and become much 
flatter than $\rho \propto r^{-2}$ at radii less than a few thousand AUs. 
Because the dust emission at (sub)millimetre wavelengths is intrinsically 
very weak, these observations however 
could not constrain the outer 
density structure of cores and the possible presence of edges. Fortunately, 
it is also possible to study these objects {\em in absorption} against a 
diffuse background (e.g. Abergel et al. 1996) in the 
mid-infrared (mid-IR).

In order to gain further independent insight into cloud core structure
prior to protostar formation, we undertook
dedicated infrared imaging observations of a broad sample of starless
dense cores with the ISOCAM camera in the mid-IR and ISOPHOT in the 
far-IR, both aboard the {\it Infrared Space Observatory} (ISO). 
The present paper summarizes the results of these ISO programs 
(see Bacmann et al. 2000 and Ward-Thompson, Andr\'e, \& Kirk 2001 for details).

We selected for study the regions which were identified by Myers and 
co-workers 
from the catalogues of Lynds and others,
and which were observed in various transitions of NH$_3$, CO and other 
molecules (Benson \& Myers 1989, and references therein). But
we selected from these a sub-sample of cores which do not contain 
IRAS sources,
on the grounds that these should be at
an earlier evolutionary stage, with the cores with IRAS sources having 
already formed protostars at their centres.
                                                            


\section{ISOCAM and ISOPHOT observations}

A sample of 24 fields were observed in the mid-IR at $\sim 7$\,$\mu$m with the camera ISOCAM (Cesarsky et al. 1996) aboard the ISO satellite (Kessler et al. 1996). 
The cores were selected so that the estimated background at 
$\sim 7$\,$\mu$m was stronger than 1\,MJy/sr so as to maximise the contrast between the absorption and the mid-IR background. A field of about 
12$\arcmin$ by 12$\arcmin$ was mapped around each core in the raster mode 
using the ISOCAM LW2 ($\lambda=[5-8$\,$\mu$m$]$) or LW6 ($\lambda=[7-8.5$\,$\mu$m$]$) filters. The raster positions were 
half-frame overlapping in the East-West direction while hardly overlapping 
in the North-South direction. In total, the integration time was 80\,s per 
raster position. The resolution of the observations (FWHM of the ISOCAM 
Point Spread Function at $\sim 7$\,$\mu $m) was 6$\arcsec$. The ISOCAM data were reduced using the CIA software package. 

Additionally, eight of these cores were also mapped in the 1.3\,mm continuum at the IRAM 30\,m telescope as well as in the C$^{18}$O(1--0) molecular line.


Eighteen cores were also observed with ISOPHOT at each of three 
wavelengths: 90, 170 \& 200\,$\mu$m.
The observations were carried out in the over-sampled mapping mode, PHT32,
in which the source is mapped by using the chopper to position the source
on the detector array at a series of positions, separated by less than the
detector resolution, intermediate between successive spacecraft pointings.
A map is built up of a series of such scans in a raster fashion.

The filters used were the C-200 filter, which has a peak wavelength response 
at $\sim 200$\,$\mu$m and bandwidth FWHM $\sim 25$\,$\mu$m, the C-160 filter, 
which
actually has a peak wavelength of $\sim 170$\,$\mu$m and bandwidth FWHM of 
$\sim 40$\,$\mu$m, and the C-90 filter, which has a peak wavelength of 
90\,$\mu$m
and FWHM $\sim 30$\,$\mu$m \cite{ref30}.
The C-200 and C-160
filters are associated with the PHT-C200 camera, which has four pixels in a 
3$\times$3 array with each pixel 45$\times$45 arcsec square.
Data reduction was carried out using the
PHOT Interactive Analysis (PIA) software version 7.1.

\section{Results}

\subsection{ISOCAM maps and comparison with millimetre continuum observations}

Out of the 24 cores observed with ISOCAM, 23 showed absorption features. For 8 of those, the absorption was strong enough 
to enable a detailed analysis. 


\begin{figure}
\plotfiddle{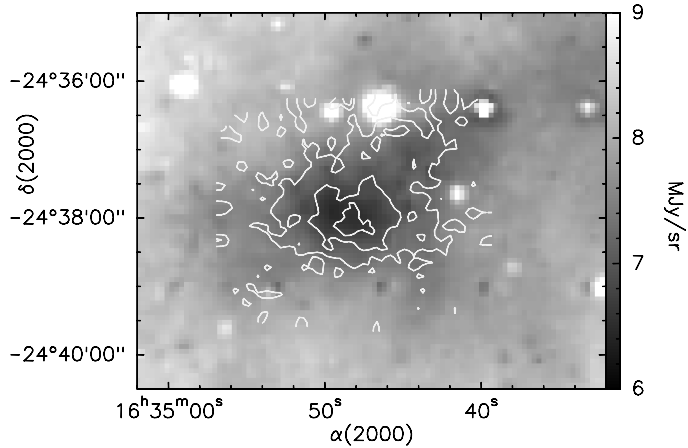}{5.9cm}{-90}{35}{35}{-150}{200}
\caption{\label{isomm}Comparison between the ISOCAM 6.75~$\mu$m image 
(greyscale in MJy/sr on the right) of the pre-stellar core L1689B (seen in 
absorption) and the 1.3\,mm continuum emission map 
(contours: 10, 30, 50\,mJy/beam).}
\end{figure}



All of them have been observed at 1.3\,mm and
present extended, weak dust continuum emission which we were able to map. For all cores, the peak of the mid-IR absorption
does correspond with a peak in the millimetre emission.
This is 
illustrated in Fig.\,1 which shows the mid-IR ISOCAM
absorption image superimposed on the 1.3\,mm continuum contours of the core L1689B, situated in one of the streamers of the Rho Ophiuchi molecular 
cloud complex. The core, seen in absorption, appears as a dark East-West 
patch towards the centre of the image, with a large-scale extension to 
the North-West.

The good agreement between the respective
images of the cores in both types of data confirms that the
features we see with ISOCAM are due to absorption by the (rather large)
column density of cold dust traced at 1.3\,mm, and not by fluctuations in
the mid-IR background.
Furthermore, the similar morphologies seen in both wavebands suggest that
the mid-IR absorption images trace the same dust as the millimetre emission maps, which justifies a detailed comparison of the column density 
profiles derived from both tracers.

\subsection{ISOPHOT maps}

Figure~2 shows ISOPHOT maps of
the same core L1689B  
at 90\,$\mu$m and 200\,$\mu$m. 
In the 200\,$\mu$m map, the core emission is clearly 
detected and some structure is seen (the 170\,$\mu$m map of L1689B, 
not shown here, is similar). 
However, it can be seen 
that at 90\,$\mu$m the core is not detected by 
ISOPHOT. All that is visible in the map is the low level extended Galactic 
cirrus emission, as expected in this low latitude region. The lack of any 
detectable emission associated with L1689B at 90\,$\mu$m shows that the dust 
which is detected at the longer wavelengths must be very cold ($T<20$\,K).

This core is typical of what we have seen throughout our ISOPHOT 
sample. Namely, we have detected all but one of the 18 cores in emission 
at the longest wavelengths, and failed to significantly detect all but 
one of the cores at 90\,$\mu$m. 


\begin{figure}
\plotfiddle{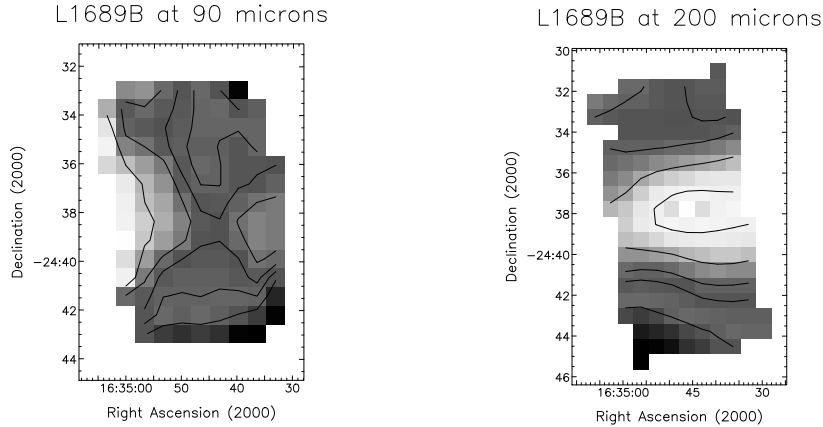}{4.7cm}{-90}{40}{40}{-160}{190}
\caption{\label{isomm}ISOPHOT 90-$\mu$m and 200-$\mu$m images of 
L1689B.}
\end{figure}

\subsection{Subtraction of the far-IR background}

The far-infrared background arises from a number of causes, including Zodiacal
emission and large-scale Galactic Plane emission. IRAS detected these
extended background emissions (Beichman et al. 1986; Jessop \& Ward-Thompson 
2000), as did the Diffuse InfraRed Background
Experiment (DIRBE) on the COBE satellite. IRAS did not cover wavelengths
beyond 100\,$\mu$m, but DIRBE extended out to 240\,$\mu$m. 
Comparison between our 
background estimates and the
COBE data indicates that the absolute calibration uncertainty of the ISOPHOT 
data is no worse than $\pm$30 per cent.

\begin{figure}
\plotfiddle{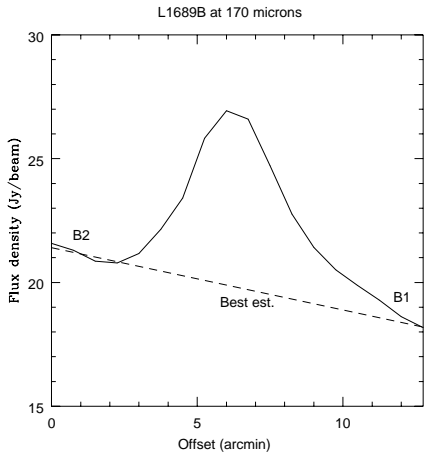}{5.cm}{0}{40}{40}{-90}{-85}
\caption{North-south flux density profile through L1689B at 170\,$\mu$m.
Notice how the core sits on a sloping background from the B2 level at the
left to the B1 level at the right.}
\end{figure}


We note that in every case the pre-stellar cores also
appear to sit on top of more extended emission that is probably
related to the larger scale cloud in which the core is embedded. 
This can be seen in Fig.~3, where we took a 1-D north-south cut through the core
L1689B at 170\,$\mu$m. 
On this plot, the background in the north is considerably higher than
that in the south. 
We estimated the value of this larger scale cloud emission by fitting 
a sloping background to the region of each image where there was extended
emission, but away from the core itself. 
After subtraction of this background emission, we measured the flux density 
from the core in a 150-arcsec diameter aperture.


\section{Analysis}

\subsection{Column density profiles from the ISOCAM absorption data}

In order to derive column density profiles from each ISOCAM mid-IR 
absorption map, we assumed a simple geometrical model. 
The dark core is embedded within its 
parent molecular cloud, and the mid-IR background arises from the outer 
parts of the molecular cloud (e.g. Bernard et al. 1993). 
This emission is thought to come from small aromatic particles distributed throughout the cloud and excited by interstellar radiation (from e.g. massive stars) up to A$_{v}$ $\sim$\,1 (Hollenbach \& Tielens 1997). 
The particles re-emit the radiation as a group of spectral lines/bands, the Unidentified Infrared Bands (UIBs), which are included in the ISOCAM filters we used. Along the line of sight to the core, we measure a background intensity $I_{back}$ arising 
from the rear part of the cloud and attenuated by the absorption from the dense core of opacity $\tau$, and a foreground intensity $I_{fore}$ arising from the front part of the molecular cloud. Hence the intensity along the line of sight can be expressed as: 
$I=I_{back}\times e^{-\tau(r)}+I_{fore}$.
The H$_2$ column density $N_{H_{2}}(r)$ at distance $r$ from the core centre is related to the opacity by: $\tau(r)=\sigma_{\lambda} N_{H_2}(r)$, where $\sigma_{\lambda}$ is the dust extinction cross-section at wavelength 
$\lambda$. We assumed $\sigma_{\lambda}$ 
to follow the Draine \& Lee (1984) model and took $\sigma_{6.75\mu m}=1.2\times 10^{-23}$\,cm$^2$ for the LW2 filter and $\sigma_{7.75\mu m}=1.35\times 10^{-23}$\,cm$^2$ for the LW6 filter.

The intensities $I_{back}$ and $I_{fore}$ are unknown quantities (assumed to be uniform over the mapped area) and were inferred from our millimetre 
observations: knowing the values of the column density in the flat central part $N_{H_{2}}^{flat}$ and in the outer part of the profile $N_{H_2}^{out}$, it is easy to estimate $I_{back}$ and
$I_{fore}$ using the above equation linking $I$ and $\tau(r)$ for $r = R_{flat}$ and $r = R_{out}$,
and then deduce the column density
from the mid-IR intensity at any position in the ISOCAM image.                            
Assuming the 1.3\,mm dust emission is optically thin, we can derive the central column density averaged over the flat inner part of the core using Eq. (2) of Bacmann et al. (2000). We assumed a single, representative dust temperature 
$T_d = 12.5$~K for all the cores 
(cf. \S ~4.3. -- see also Ward-Thompson et al. 2001).
Considering the additional uncertainty on the millimetre
dust mass opacity in dense cores,
which we took equal to $\kappa_{1.3mm} = 0.005$~cm$^{2}$g$^{-1}$
(see, e.g., Andr\'e, Ward-Thompson, \& Motte 1996),
the estimates of $N_{H_{2}}^{flat}$ calculated are
uncertain by a factor of $\sim 2$ on either side of their nominal values.

The second value of the column density we can use to calibrate the ISOCAM
column density profile is given by our C$^{18}$O(1-0)
line observations of the outer parts of the cores. Assuming optically thin emission, Local Thermodynamical Equilibrium (LTE) and a gas kinetic temperature of $\sim 10$\,K, we use Eq.~(3) of Bacmann et al. (2000) to estimate the C$^{18}$O column density in the outer parts of the core.
Given the uncertainties (a factor of $\sim 2$ on either side of the nominal values) in the determination of the column densities, in addition to `best' values for $I_{back}$ and $I_{fore}$, we obtain maximum and minimum values which define a range of profiles compatible with the data. For L1689B, we find $I_{back}=3.8$\,MJy/sr and $I_{fore}=4.5$\,MJy/sr. $I_{fore}$ includes a zodiacal light contribution of $I_{zodi}=4.3$\,MJy/sr.

\begin{figure}
\plotfiddle{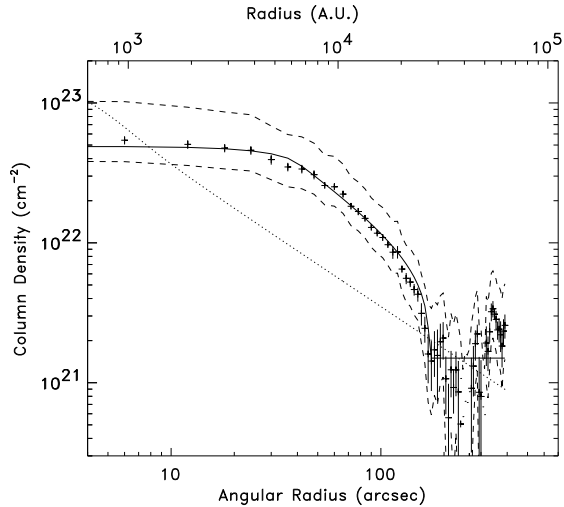}{6.5cm}{90}{40}{40}{140}{-40}
\caption{\label{profilsouth}Column density profile of the core L1689B in the 
North-South direction (crosses). The dashed lines limit the range of profiles 
compatible with the data. The dotted line represents the singular isothermal 
sphere convolved with the ISOCAM PSF. The solid line is a parameterised Bonnor-Ebert-like model.}
\end{figure}


In order to characterize the radial structure of each core and reduce
the influence of spatial fluctuations in the background and foreground,
we first derived mean radial intensity profiles by
averaging the mid-IR emission in the ISOCAM images according to the
apparent morphology of the core. 
An example of a 
column density profile obtained using the method described above is shown in Fig.\,4 as crosses whose vertical sizes represent the error bars on the data points. The dashed curves lying above and below the `best' 
profiles limit the range of column density profiles that are consistent with 
the mid-IR absorption data, given the permitted range of values for 
$I_{back}$ and $I_{fore}$ in each case. The dotted line represents the  
radial profile of a Singular Isothermal Sphere (Shu 1977) at 10\,K, 
convolved with the ISOCAM PSF (see discussion in Sect~5.1). 


All the profiles of the cores studied here present a flattening in their inner regions. This confirms previous results obtained from (sub)millimetre continuum measurements (Ward-Thompson et al. 1994, Andr\'e et al. 1996). The outer shapes and slopes of the profiles on the other hand can be divided into two broad categories. For roughly half of the cores, the slope of the profiles follows approximately a $N_{H_{2}}\propto r^{-1}$ power-law up to the boundary of the map. For the remaining cores, the column density profile steepens with increasing projected radius until it reaches a plateau where the dense core merges with the ambient molecular cloud. 
In three cases even (including L1689B pictured in Fig.\,4), the profile presents an ``edge'', where
we define an edge as a profile steeper than $N_{H_{2}}\propto r^{-1}$. This feature is new and has interesting implications. The presence of a sharp edge at $R\sim 130$\,arcsec in the column density profile of L1689B (cf. Fig.\,4) is consistent with the core being unresolved by ISOPHOT in the North-South direction (cf. Fig.~2 and Fig.~3). In Sect.~5, we compare the obtained profiles with the predictions of a few models of core structure.         

\subsection{Colour temperatures from the ISOPHOT data}

We measured the colour temperature variation across each
core by first subtracting the far-IR background level 
that we had measured in each image according to our best estimate,
as described in \S ~3.3. We then ratio-ed the 
background-subtracted images at 170 and 200\,$\mu$m. We converted this to
a colour temperature using the usual
black-body assumptions -- namely that:

\[ (F_{\nu(1)}/F_{\nu(2)}) = \frac{\nu_{(1)}^3(e^{[h\nu_{(2)}/kT]} - 1)}
{\nu_{(2)}^3(e^{[h\nu_{(1)}/kT]} - 1)}, \]

\noindent
where the frequencies $\nu_{(1)}$ and $\nu_{(2)}$ correspond to
wavelengths of 200 and 170~$\mu$m respectively, and $F_{\nu(1)}$ and
$F_{\nu(2)}$ are the flux densities at each of these frequencies.
$T$ is the dust temperature, and
other symbols take their usual meanings. This is
the simplest set of assumptions that can be made. It assumes
that all of the dust in a given pixel
is at a single temperature, $T$, and that it is
exactly the same dust which is emitting at both wavelengths. It also
assumes that the dust is not self absorbing and consequently has zero
optical depth. 
Using these assumptions we constructed a series of colour temperature
maps.

\begin{figure}
\plotfiddle{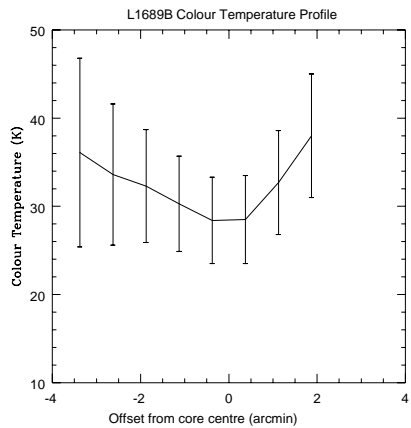}{5.cm}{0}{38}{38}{-100}{-70}
\caption{Colour temperature cut through L1689B along its major  
(i.e., east-west) axis, based on grey-body fits
to the ratio of the 200\,$\mu$m and 170\,$\mu$m data. 
(Note that this colour temperature overestimates the physical 
temperature.)}
\end{figure}

Figure~5 shows a colour temperature cut through the pre-stellar core
L1689B. This has been calculated by first removing our best estimate of
the background flux density from each of the two input images, and then
ratio-ing them. A lookup table was created using the optically thin
grey-body assumption with $\beta$=2. The values in the ratio image are
then converted into formal colour temperatures (see Ward-Thompson et al.
2001 for a full description). Figure~5 shows tentative evidence that
L1689B may be colder in its centre than it is towards the edge.
Ward-Thompson et al. (2001) found that 8 out of 18 cores were definitely
cooler at their centres, with a further 2 cores (including L1689B) showing
some evidence for the same trend. This appears to confirm the pre-stellar
nature of these cores, with no central heating source, and the source of
energy being the external inter-stellar radiation field.                        

\subsection{Spectral Energy Distributions}

We can measure the flux densities of our pre-stellar cores
at 200 \& 170\,$\mu$m and estimate upper limits to the flux density at 
90\,$\mu$m,
and hence construct the spectral energy 
distribution (SED) of each source.
Figure~6 shows a plot of 
Log(S$_\nu$) against Log($\nu$) for the 
pre-stellar core L1689B where the flux densities have been measured
in its central region.
The plot shows the ISOPHOT 90-$\mu$m upper limit and 170 \& 200-$\mu$m 
detections, as well as the JCMT 850-$\mu$m and IRAM 1.3-mm detections.

We have fitted the source with a modified 
black-body function (often referred to as a grey-body)
of the form:
$S_{\nu} = \Omega\, B_{\nu} (T_{\rm d})\, (1 - {\rm 
exp}[-\tau_\nu] )$, 
where $B_{\nu} (T_{\rm d})$\ is the blackbody function, 
$\Omega$ is the solid angle of
the source, $\tau_{\nu}$ is the 
optical depth (=$[\nu/\nu_c]^\beta$),
$\nu_c$ is the critical frequency at which the optical depth is 1,
and $\beta$ is the dust emissivity index. We show one
such fit as a 
solid line on Fig.\,6, which has $T=11$\,K, $\beta$=2,
typical of our sources -- all are $<$\,20\,K.
Note that the temperature derived from fitting the full SED is less than
that which was derived from simply ratio-ing the 170 and 200\,$\mu$m data
in \S ~4.2 above. When the data at other wavelengths are taken into 
account, chiefly the 90\,$\mu$m upper limit, the best fit temperature 
decreases from 28\,K 
to 11\,K.

\begin{figure}
\plotfiddle{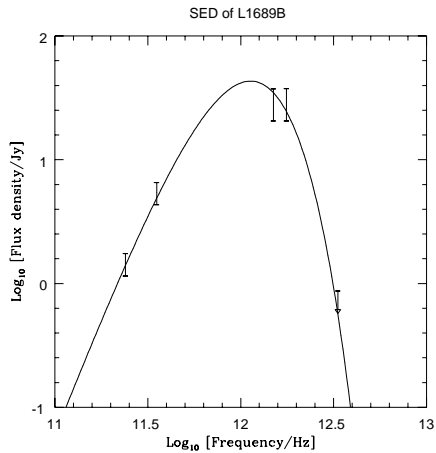}{4.7cm}{0}{40}{40}{-120}{-86}
\caption{Spectral energy distribution of the pre-stellar core L1689B.}
\end{figure}
Integrating under the curve in Fig.\,6, 
we estimate the output luminosity radiated by L1689B to be 
$\sim 0.3$~L$_\odot$. 
Estimating the local inter-stellar radiation field (ISRF) incident on L1689B 
from the measured mid-IR and far-IR backgrounds 
($I_{back}^{7\mu m} \sim 4$\,MJy/sr and 
$I_{back}^{200\mu m} \sim 115$\,MJy/sr -- cf. \S ~3 and \S ~4 above) 
and the spectral shape advocated by Mathis, Mezger, \& Panagia (1983), 
the amount of external heating received by the core can be calculated.
In this way, we infer that the input luminosity absorbed by L1689B is  
$\sim 0.4$~L$_\odot$ with a factor of $\sim 2 $ uncertainty. 
Ward-Thompson et al. (2001) find a similar result for a number of
pre-stellar cores. Hence these cores are consistent with having no central
heating source.

\section{Confrontation with theoretical models}

\subsection{Singular isothermal sphere (SIS) and logotrope}

In the widely used `standard' paradigm of isolated star formation (e.g. Shu et al. 1987), the SIS provides the initial conditions of gravitational collapse 
and could a priori be representative of the most advanced pre-stellar cores. 
It results from the equilibrium between gravitation and thermal pressure at 
all points. Such a self-gravitating sphere has a power-law density profile 
following  $\rho(r) = (a^2/2\pi G)r^{-2} $ , where $a$ is the 
effective isothermal sound speed (ie a {\em column density} profile 
in $N_{H_{2}}\propto r^{-1}$).
The column density profile of the SIS at 10\,K (the assumed gas temperature of 
our cores) however fails to reproduce the flattening in the inner part of the 
profile due to its central singularity, and cannot account for the edges in 
the outer part either (see Fig.~4).
 


McLaughlin \& Pudritz (1996) have advocated
a special non-isothermal equation of state for star-forming molecular clouds
of the `logotropic' form $P/P_c = 1 + A\, $ln$(\rho/\rho_c)$, where
$P_c$ and $\rho_c$ denote the central values of the pressure and density,
respectively, and $A$ is a constant whose realistic value is
$A \approx 0.2$. Such objects follow Larson's phenomenological relations (Larson 1981) for molecular clouds. When we confront the column density
profiles inferred from our ISOCAM data to the profiles of finite-sized
logotropes at a gas temperature of T=10~K,
bounded by a surface pressure $P_{s}$,
we find that the logotrope cannot match the observed profiles: beyond the flat inner region, the ISOCAM profile approaches a slope in  $N_{H_2}\propto r^{-1}$ while the logotrope is substantially flatter than $N_{H_2}\propto r^{-1}$ 
(Bacmann et al. 2000).

\subsection{Ambipolar diffusion models of magnetically supported cores}

A natural way to account for the flat inner part and sharp outer edges is to invoke models of magnetically supported cores undergoing ambipolar diffusion (e.g. Mouschovias 1991, Ciolek \& Mouschovias 1995 -- hereafter CM95. See also 
Ciolek \& Basu, this volume). According to these models, the core first 
contracts quasi-statically through ambipolar diffusion. During this 
magnetically
{\it subcritical} phase, the central mass-to-flux ratio $(M/\Phi)_{cent}$
increases until it reaches the critical value $(M/\Phi)^{crit}_{cent} = (1/2\pi )G^{-1/2}$ (e.g. Basu \& Mouschovias 1995). At this point, the inner central core becomes magnetically {\it supercritical}
and the evolution speeds up. The supercritical core collapses dynamically
inward, while the outer (subcritical) envelope is still efficiently supported
by the magnetic field and remains essentially ``held in place''. As a result,
a (very) steep density profile or `edge' develops at the boundary of the
supercritical core.

We have quantitatively compared the inferred profiles with the models calculated by CM95 at various stages of evolution. We find a good agreement between the data and the 
core model~B$_{UV}$ of CM95 near time $t_{2}$, i.e., the time at which the
central density has increased by a factor $10^2$ (and the central
column density by a factor 10) compared to its initial value. This is shown
in Fig.\,7 for another pre-stellar core L1544 similar to L1689B
(e.g. Tafalla, this volume).

There are however two main problems with these ambipolar diffusion 
models: first, the models predict that the cores should have a highly 
flattened disk-like structure, whereas in our ISOCAM images a majority of 
them have a filamentary appearance on large scales. 
Second, the value of the magnetic field needed to account for the sharp 
edges in the profiles is high, typically 
of order $\sim 35$\,$\mu$G in the case of L1544. 
This is higher than the 
Zeeman measurements of magnetic field intensities towards dark clouds 
which indicate typical field 
strengths (or upper limits) $\sim$\,10\,$\mu$G 
(e.g. Crutcher 1999, Crutcher, this volume). For example, Crutcher \& Troland (2000) have measured a magnetic field of 11\,$\mu$G towards L1544, in agreement with the model developed by Ciolek \& Basu (2000) to account for the infall velocities in the core (Williams et al. 1999). 
Submillimetre polarimetric observations of the magnetic field in 3 cores
(Ward-Thompson et al. 2000) also tend to favour those models with a low
initial magnetic energy density. However, the stength of the magnetic field in these models is too low to account to the sharp edges we find for L1544.                                                                         

These results suggest that turbulence might play a role in the outer parts 
of the cores. It could indeed provide additional support against collapse 
and hence allow the static magnetic field to be lower with regards to the 
situation where there is no turbulence, and also account for the filamentary 
shapes of the cores. 
Note, however, that the C$^{18}$O(1-0) lines we have measured 
towards the outer regions of the cores are relatively 
narrow ($\Delta V_{FWHM}\sim 0.4$\,km/s).

\begin{figure}
\plotfiddle{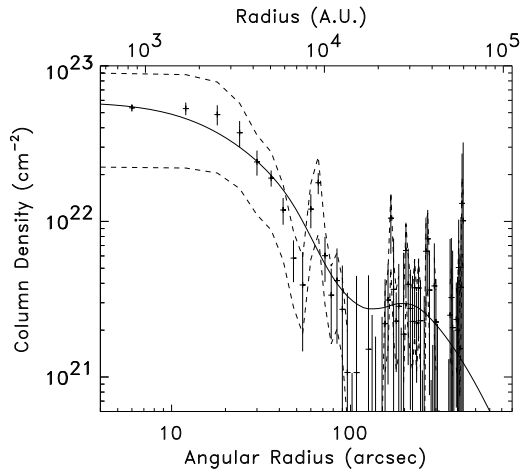}{5.3cm}{90}{35}{35}{160}{-15}
\caption{\label{ciolekmodel}Comparison of the column density profile of 
L1544 and magnetic model CM95 (solid line).}
\end{figure}






\section{Summary and implications}

We observed a sample of candidate pre-stellar cores with the mid-infrared camera ISOCAM and in the far-infrared with ISOPHOT. These measurements are consistent with the cores being starless and enabled us to derive their temperatures.
Using column density measurements from millimetre continuum and millimetre line observations, we have derived the radial column density profiles of a subsample of objects. 
A new feature 
is that the outer profile of three of our cores present sharp edges (corresponding to radial column density
profiles that drop steeper than $N_{H_{2}} \propto r^{-2}$).  An important implication is that the dense cores are decoupled from their parent molecular clouds and that the mass reservoir available for star formation in these cores is finite, supporting the idea that stellar
masses are partly determined at the pre-stellar stage (cf. Motte et al., 
this volume).
Ambipolar diffusion models of magnetically supported cores can reproduce the flat inner parts and the sharp edges of the profiles, although they require  stronger background magnetic fields ($\sim$\,30\,$\mu$G) than those usually observed in these objects (Crutcher, this volume). 
The addition of a turbulent field to these models might soften the constraints 
on the static magnetic field and provide an explanation for the filamentary 
shapes observed for the cores.

%

\end{document}